\documentclass[referee]{raa}           
\usepackage{graphicx,times}
\usepackage{natbib}
\usepackage{amssymb,amsmath}
\usepackage{epsfig}
\usepackage{pdfpages}

\bibpunct{(}{)}{;}{a}{}{,}
\usepackage{hyperref}
\hypersetup{pdftitle = The title of my PDF, pdfauthor = My name, pdfsubject= The subject, pdfkeywords = keyword1 keyword2 keyword3} 
\hypersetup{colorlinks = true, linkcolor = green, anchorcolor = red, citecolor = blue, filecolor = red, pagecolor = red, urlcolor = red}

\begin{document}

\title{Full-disk Synoptic Observations of the Chromosphere Using H$_{\alpha}$ Telescope at the Kodaikanal Observatory}

 \volnopage{ {\bf 2015} Vol.\ {\bf X} No. {\bf XX}, 000--000}
   \setcounter{page}{1}

   \author{Belur Ravindra\inst{1}, Kesavan Prabhu\inst{1}, Komandur Elayaveilli Rangarajan\inst{1}, 
Bagare P. Shekar\inst{1}, Singh Jagdev\inst{1},
      Kemkar Madan Mohan\inst{1}, Paul Lancelot\inst{1}, 
Koyipurathu Chellappan Thulasidharen\inst{1}, F. Gabriel\inst{1} and Raju Selvendran\inst{1}}

   \institute{Indian Institute of Astrophysics,
              IInd Block, Koramangala,
              Bengaluru - 560 034;
              {\it ravindra@iiap.res.in}\\
\vs \no
   {\small Received  --; accepted --}
}

\abstract{This paper reports on the installation and observations of a new solar telescope installed on 7th October, 2014 at the Kodaikanal Observatory. The telescope is a refractive 
type equipped with a tunable Lyot H$_{\alpha}$ filter.  A CCD camera of 
2k$\times$2k size makes the image of the Sun with a pixel size of 
1.21$^{\prime\prime}$ pixel$^{-1}$ with a full field-of-view of 41$^{\prime}$. The telescope 
is equipped with a guiding system which keeps the image of the Sun within a
few pixels throughout the observations. The FWHM of the Lyot filter is 
0.4\AA~and the filter is motorized, capable of scanning the H$_{\alpha}$ line profile at a smaller step size of 0.01\AA.  Partial-disk imaging covering about 10$^{\prime}$, is 
also possible with the help of a relay lens kept in front of the CCD camera. In this paper, we report the detailed specifications of the telescope, filter unit, its installation, observations
and the procedures we have followed to calibrate and align the data. We also present 
preliminary results with this new full-disk telescope.
}

   \authorrunning{Ravindra et~al.}            %author_head in even pages
   \titlerunning{Full-disk synoptic observations of the chromosphere}  % title_head in odd pages
   \maketitle

%________________________________________________ sections below
%
\section{Introduction}           %% first-level sections will be auto-capitalized
\label{sect:intro}

Apart from white-light observations of the Sun, the data recorded in
H$_{\alpha}$ wavelength have a long history. The chromosphere was seen in red and pink 
color during the total solar eclipse at the beginning of the 18th century. 
During the beginning of the 19th century, observers reported the brilliant, glowing 
prominences in red color. Later, several others including Jules Jansen observed 
the spectra of the prominences during the total solar eclipse of 1868. In 1889 Hale 
invented the spectroheliograph to map solar prominences above the solar limb. 
Apart from Mt. Wilson, several other 
observatories around the world  started the observations of the Sun in different 
wavelengths using the spectroheliograph. Kodaikanal observatory started its Ca-K 
and H$_{\alpha}$ observations in 1905. These observations were made using the 
spectroheliograph and siderostat as a light feeding system \citep{bappu1967}
%(Bappu~\citep{bappu1967}). 

In the next few years, several other observatories around the world  started the observations 
in H$_\alpha$ and Ca-K  (Meudon observatory in France, Sunspot in New Mexico, Nainital 
and Udaipur Solar Observatories in India, Big Bear Solar Observatory in California, 
Hida observatory in Japan and many more). The invention of birefringent filters 
by \cite{ohman1938} in Sweden and \cite{Lyot1944} in France started a
new era in observing the Sun using narrow band filters that were able to isolate the spectral line of the desired wavelength. From 1950 to the end of the 20th century,
 Messers B. Halle company 
in Germany used to make those filters and at the same time Carl Zeiss also made a few such 
filters with 0.25 to 0.5~\AA~passband. The set up could also scan the spectral line profile
and made observations of the Sun at different heights in the solar atmosphere. Several
observatories around the world  utilized these filters and made solar observations
in Ca-K and H$_\alpha$.  It became possible to take images of the Sun at a faster 
rate with these filters as compared to taking the images using the exit slit of the
spectroheliograph. The introduction of CCD cameras in the 90's further enhanced the image
acquisition rate with high contrast. 

During the beginning of the 21$^{st}$ century, observatories around the world started 
a global H$_{\alpha}$ network with a nearly 24 hour coverage of the Sun. In 2009, 
the Global Oscillation Network Group  \citep[GONG;]{Hill2009} added  H$_{\alpha}$ to 
its regular observations along with dopplergrams, magnetograms, and pseudo continuum images of the Sun.  Apart from these, several other observatories around the world have their own program of H$_{\alpha}$ observations, for e.g., Improved Solar Optical Observations Network \citep[ISOON;]
{Neidig98} which belongs to the US air force, The {\it Chromospheric 
Telescope} \citep[ChroTel;]{bethge2011} installed at the Observatorio 
del Teide, Tenerife makes full-disk observations of the solar chromosphere in Ca II K, 
H$_{\alpha}$ and He I 10830~\AA. The Aryabhatta Institute 
for Observational Sciences \citep{Verma97} has set up an H$_{\alpha}$ observational facility. Similarly, the Udaipur Solar Observatory started the H$_{\alpha}$ 
observations since 1970. Thus, ground-based telescopes established at 
different longitudes around the globe are very useful to monitor solar activity with 
24¬hr coverage.

Kodaikanal observatory has a history of solar observations made in white-light, Ca-K and H$_{\alpha}$ wavelengths since 1905. The H$_{\alpha}$ observations continued till 2000. Due to the unavailability of the photographic plates, the observations were discontinued, although
the same siderostat
was used for the Ca-K observations with a CCD camera. Following the installation of the `twin telescope' at the Kodaikanal observatory  \citep{jagdev2012}, white-light
observations were carried out simultaneously with Ca-K. To continue 
solar chromospheric observations in H$_{\alpha}$, we have recently installed a new telescope 
at Kodaikanal observatory. In this paper we describe the details of the telescope, control system, the guider unit to stabilize the telescope, the filter unit used to isolate the H$_{\alpha}$ spectral line, scanning of the H$_{\alpha}$ line profile to observe the different layers of the Sun etc.  We also show some of the observational results obtained with this telescope and outline some of the future plans.

\section{Scientific Outlook}
H$_{\alpha}$ telescopes were used more than 100 years for solar observations. Several  results from various data sets were published over the years.  H$_{\alpha}$ images 
are used, either as supplementary data \citep{rohan2015, vemareddy2012}, or as a stand-alone 
data set \citep{sri1991, venkat2008, wang1998, makarov1989}. Below we illustrate
a few scientific goals that can be addressed with the newly installed H$_{\alpha}$ telescope 
at the Kodaikanal observatory. They are only representative but not exhaustive. 

Solar filaments are chromospheric and coronal features, appear as large, dark, thread like
structures with certain thickness on the solar disk.
Typically, quiet Sun filaments are larger than active region filaments. 
The largest ones cover more than half the Sun's diameter (e.g., on 11 February, 2015). The observations made in H$_{\alpha}$ provide information on the large scale mass motion inside the filament during its eruption and it is important for modeling solar filament eruption, hence observations made in this wavelength are important for understanding the onset of solar flares and coronal mass ejections.

Solar filaments are complicated structures, yet very important features observed in the 
the corona. Several types of oscillations in filaments are observed in the corona that are 
excited during a solar flare. Solar flares can excite fast MHD shock waves such as, Moreton 
waves in the chromosphere \citep{okamoto04, gilbert08} and EIT waves in the corona 
\citep{eto02}. These waves propagate with velocities in the range 400-1500 km~s$^{-1}$ and can excite the oscillations in prominences which dampens over 
time. The oscillatory properties of solar filaments, associated with flaring active regions, 
are significant for understanding their stability and the relevant conditions leading to 
their eruption. To detect the above phenomenon in the chromosphere, we would require fast
cadence full-disk images taken in the H$_\alpha$ line core during flares and filament eruptions.  This would help us to identify the cause of the large amplitude oscillations of the filaments.

In 2-D model of a flare, the energy release occurs in the current sheet and the 
coronal magnetic reconnection rate can be computed by summing the normal component 
of the photospheric magnetic field at the locations of flare ribbons 
in the lower solar atmosphere \citep{priest86}. 
The flare ribbons are believed to be the maps of the footpoints of the magnetic field 
lines reconnected in the corona. From the assumption of conservation of magnetic flux, 
one can compute the total magnetic reconnection flux by using the observations of 
flare ribbons in the lower atmosphere and the corresponding photospheric magnetograms.
While this technique is used to compute the magnetic flux involved in the flare, in most 
of the cases the TRACE images taken in C IV 1600~\AA~are used 
\citep{long2007} for finding locations of the flare ribbons and newly reconnected regions. The C IV 1600~\AA~wavelength shows both the chromospheric and 
photospheric information. The line core of the H$_{\alpha}$ is formed in the upper 
chromosphere and the wings are formed closer to the photosphere. One can make the comparison
of the total magnetic reconnection flux involved in the 
flare using the flare ribbons observed in H$_{\alpha}$ and C~IV~1600~\AA~wavelength. 
This will provide an important information about total reconnected magnetic flux involved in 
the flare which can be compared with the magnetic flux carried away by the magnetic 
clouds \citep{qiu2007}.

\section{The Telescope}
\subsection{Optics}
The main telescope installed at the Kodaikanal observatory has a 20.06-cm doublet 
lens as an objective which makes an f/7.9 beam. This beam
is collimated by another doublet having a diameter of 5.2-cm and a focal length of 18.94~cm. 
The light from the collimator is refocused by a re-imaging lens that has an effective 
focal length of 25.6~cm.

\begin{figure}[!h]
\begin{center}
   \includegraphics[width=0.5\textwidth]{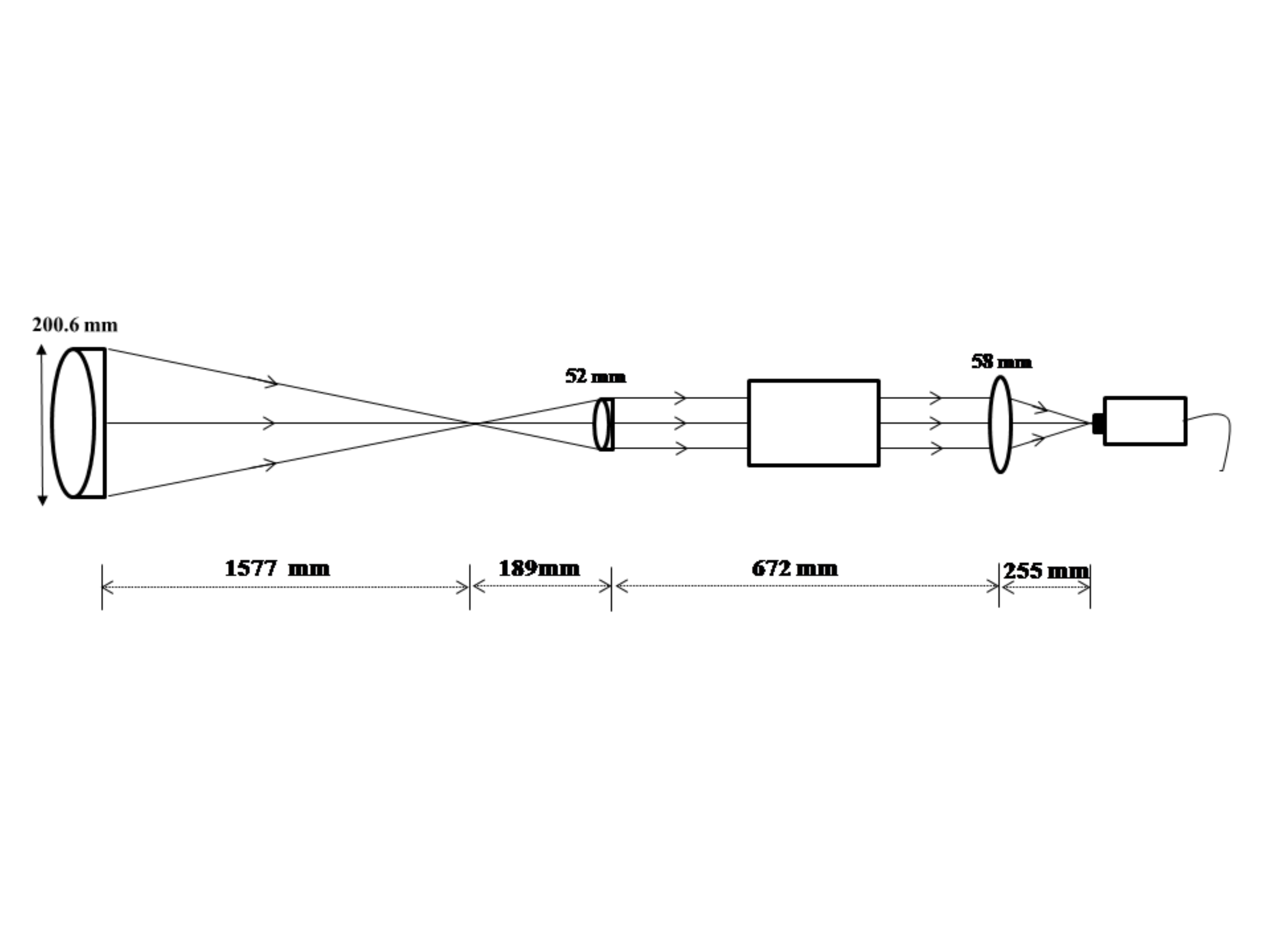}
   \end{center}
   \caption{\small A schematic layout of the optical configuration of the H$_{\alpha}$ main telescope.}
\label{fig:1}
\end{figure}

The final image size is about 2.1~cm for a 32$^{\prime}$ field-of-view (FOV) and about 2.7~cm 
for a 41$^{\prime}$ FOV. The H$_{\alpha}$ filter  is kept in the collimated beam. A 
combination of lenses with an effective focal length of 16.52~cm is placed after the re-imaging
lens to obtain the magnified view of the Sun. With this set-up one-quarter of the Sun's image can be obtained.  Both the full-disk and partial-disk mode can
be used to study some of the events such as prominence eruptions on the limb, solar flares, 
filaments, solar macro spicules etc.  The final combination of 
lens is an optional arrangement and need not be used all the time. 
The schematic design of the telescope is shown in Figure~\ref{fig:1} with dimensions of the 
lenses.

There is a guider telescope fitted to the side of the main telescope to correct for 
any image motion during observations. The guide telescope is made up of a 10-cm objective 
lens with an effective focal length of 174.3~cm. The lens is cut in to 4 quadrants and each quadrant makes a solar image with partial images overlapping each other. The region 
between the overlapped Sun's image is dark. A CCD camera of 795${\times}$596 pixels is 
kept in the dark
portion of the overlapped image. Any shift of the Sun's image due to tracking or shaking 
will produce an error signal which is amplified and given as feed back to 
the telescope. The guider telescope helps to keep the image in the main telescope CCD 
camera to within a few pixels throughout the observation.  

Apart from these, there is one more small telescope which serves as a finding tube. It 
has a 3~cm diameter small convex lens with a focal length of 252.5~cm. It forms an image 
of about 2.5~cm size on a ground glass prism. A yellow colored filter with a broad pass band (5000-6000~\AA) is kept in front of the objective lens to get a colored image of the Sun 
and also to reduce the heat load on the system. This arrangement makes it easy for the 
initial alignment of the telescope towards the Sun. 

The front of the telescope objective is covered with a stepper motor controlled shutter 
to open and close the lid which prevents dust from entering the telescope system 
when not in use. Similarly, one more automated shutter is kept in front of the guider 
telescope as well. 

\begin{figure}[h]
\begin{center}
   \includegraphics[width=0.9\textwidth]{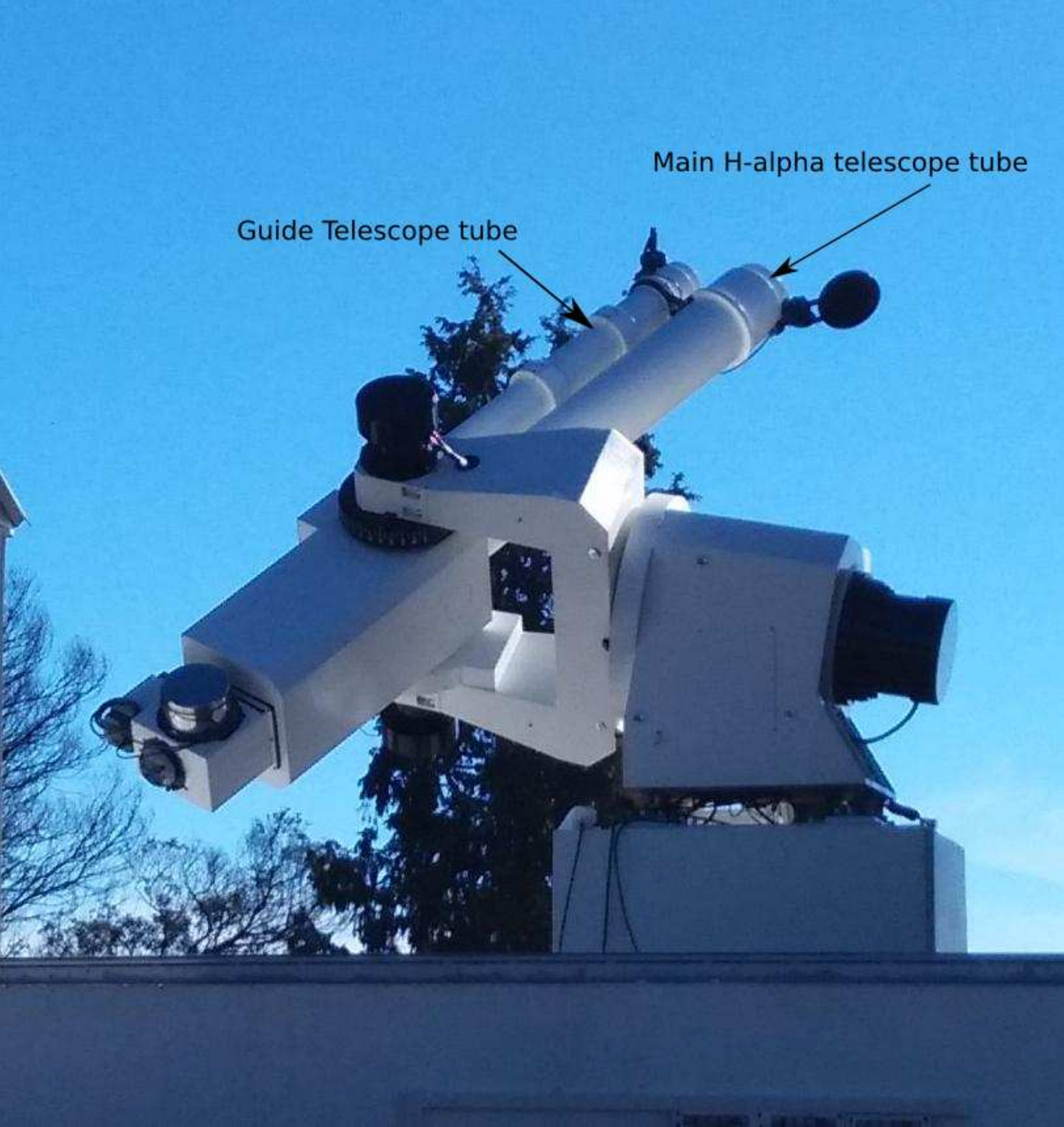}
   \end{center}
   \caption{{\small A picture of the H$_{\alpha}$ telescope installed at the Kodaikanal Observatory.}}
\label{fig:2}
\end{figure}

The overall telescope tube length is about 3.2~m. The whole assembly of 
the telescope including the guiding system is assembled on a German 
equatorial mount. The mount is installed on a reinforced 
concrete pier with a steel plate on the top of the foundation. The tracking system 
is controlled by a motor which in turn gets
the signals from  RA and DEC encoders. The picture of the telescope is shown 
in Figure~\ref{fig:2}. The main and guider telescopes are shown with red arrows 
in the same Figure. 

\subsection{The H$\alpha$ Filter Unit}
The H$_{\alpha}$ spectral line is isolated by a tunable Lyot 
filter. The Lyot filter was made using birefringent crystals which 
split the incoming ray into the ordinary and extraordinary rays. The emergent rays interfere 
at the polarizer which produces a series of interference patterns. A series of 
such retarders in combination with polarizers produce a filter. The transmission function of 
each stage of a polarizer and retarder is different. The thickest retarder decides the final bandwidth of the transmitted peak and the thinnest one determines the 
distance between successive transmission peaks, which is also called the Free Spectral 
Range (FSR). The Lyot filter used at the Kodaikanal H$_{\alpha}$ telescope has seven stages 
of crystals and the thickest crystal width is about 28~mm.

A stepper motor has been attached to the filter unit and the filter can be tuned to any 
position on the line profile with an accuracy better than 10~m\AA. The filter unit can 
scan the line profile about 4~\AA~away from line center on both wings which is 
computer controlled. The shifting of the center of the filter to any desired position on the 
line can be achieved in about 1~sec. The FSR of the filter is 
51.2~\AA~and the filter band width is 0.4~\AA~obtained from the laboratory measurements. 

The birefringent crystals are very sensitive to temperature. The change in the retardation,
shifts the passband away from the desired wavelength of interest. Hence it is very 
important to maintain a uniform temperature within the filter unit. Typically, 
quartz produces a greater shift in wavelength than calcite. The one used in the Kodaikanal 
H$_{\alpha}$ telescope has two layers of heating units. The entire filter unit is 
maintained at a temperature of 42$^{\circ}$~C with a stability of 0.01$^{\circ}$~C using 
a heat sensor and feedback unit. The filter works fine even if the ambient temperature 
varies from --20$^{\circ}$ to 35$^{\circ}$~C. 

The acceptance angle of the filter is large (about 2$^{\circ}$) over which the shift in the
wavelength is less than 0.1~\AA. The overall size of the filter unit is 0.5~m in length, 0.2~meter width, 0.2~meter height and weighing about 35~kg. The clear aperture of the filter
unit is about 3.6~cm. The crystal unit along with the polarizers is immersed in  
silicone oil. Silicone oil has several good properties which are very useful for optical 
elements when we used as a one single unit. Some of the properties of silicone oil are; 
it is insoluble in water, it has a low viscosity,  
it is colorless, it has a very low freezing point, very high boiling point, it has good thermal 
conductivity, it is a poor conductor of electricity, and most importantly the refractive index 
of the oil is same as the other crystals in the unit thereby reducing the back reflection 
from each surfaces. 

A pre-filter from Andover Corporation, having a width of 30~\AA~and centered on H$_{\alpha}$,
is kept in front of the Lyot filter which isolates the spectral line of interest The overall 
transmittance of the H$_{\alpha}$ filter is larger than 5\%.  

\subsection{Telescope Mechanical Structure and Drive Units}
The equatorial configuration of the telescope mount allows tracking the Sun 
by rotating about a single axis. The mount is made of cast iron, and it is stable and rigid.
It does not require any additional counterweight for balancing. The equatorial 
fork mount places the polar axis at an angle of 10$^{\circ}$, which corresponds to the 
latitude of the location. Further fine adjustment in the alignment was made in the 
telescope pedestal for latitude and azimuth. Finally, the telescope was aligned to the polar 
axis with an accuracy better than 0.5$^{\circ}$.

The telescope is coupled to the direct drive motor system which utilizes permanent magnet 
DC torque motors. The torque rating of the motors is 50 N-m and 25 N-m in the two axes to control 
the motions in RA and DEC directions, respectively. The RA axis operates at three speeds of 
2$^{\circ}$s$^{-1}$, 4$^{\circ}$ min$^{-1}$ and 5$^{\prime\prime}$ s$^{-1}$. The primary advantage of a direct drive system is that it has a hollow shaft and it provides a continuous 
torque with high positional stability and bi-directional repeatability. This property 
of the direct drive motors eliminates mechanical transmission elements such as gears, belts, friction contact etc. thereby providing positioning, pointing, and tracking  with good accuracy.

An internal brake system is installed for both the axes which ensures the safety of the 
telescope during operation and also takes care in the event of a power failure. A gear 
and pinion assembly controls the acceleration of the load during power failure and transmits 
the motion to the motor to engage the brake. 

\subsection{Encoder Alignment on RA and DEC Axis}
The absolute incremental encoders are used to track the current position of the telescope by getting the information about the rotational motion of the shaft. The advantage is that 
the absolute encoders store the information about the position even when the power is off. 
The encoder ring was aligned with high accuracy against the encoder head. This provides the precise information about the position of the telescope in the RA and DEC axes with an accuracy 
of 1$^{\prime\prime}$.

\subsection{Telescope Operation}
The telescope has been operational since October, 2014. Typically the telescope is operated in 
full-disk mode and the images are taken in the H$_{\alpha}$ line center. The H$_{\alpha}$ 
line center images are saved every minute. This is 
to maintain the data rate at par with other observatories around the world and at the same 
time manage data storage. The flats and darks are taken once a day.  During a flare or 
filament eruption, the images are taken at an interval of 30-sec. Occasionally, we scan the 
H$_{\alpha}$ line profile at 6 positions -- one in the line center, 2 each on the red and 
blue wings (0.4~\AA~and 0.8~\AA~from line center) and finally at +1.2~\AA~away from line center, 
which represents a quasi-continuum point. Each of these images are taken with a cadence of 
10 sec.

\subsection{Detector System}
We employ a CCD camera with a grade-I chip supplied by Andor technology. 
That has  2048$\times$2048 pixels. The 
pixel size is 13.5$\mu$m which is about 1.21$^{\prime\prime}$ per pixel for the 
H$_{\alpha}$ full-disk image and 0.48$^{\prime\prime}$ for partial disk.The full well capacity 
is 8$\times$10$^{4}$ electrons. The CCD is back-illuminated whose quantum efficiency is 
about 91\% at 656.3~nm 
H$_{\alpha}$ wavelength band. The CCD camera has a 16-bit digitization with a readout rate of 
1MHz. The CCD is 
cooled to a low temperature by Peltier cooling for low dark current. During observations, the 
CCD is cooled to --40$^{\circ}$~C. At this temperature the dark current is about 
1~electron/pixel/sec. The readout noise is less than 0.1\%. The normal exposure time 
used for the full-disk images is about 60-70~ms and for the partial-disk observations it 
is about 0.4-0.5~s. It has a mechanical shutter and below 20~ms exposure, the shutter pattern
appears and the counts are less than 2${\times}$10$^{4}$.  
The software program was written in Borland C++ language for the data acquisition and 
also to scan the line profile using the Lyot filter.

\section{The Data Calibration}
The data obtained from the telescope has to be calibrated before using 
it for further scientific analysis. In the following subsections we describe the details 
of the calibration procedure. 

\subsection{Flat Fielding}
There are several methods given in the literature for the flat-fielding of 
solar images. One of the widely used method is the one proposed by 
\cite{kuhn91}. In this method, gain tables are generated by using 
shifted solar images. We obtained 9 shifted images and computed the gain table. However, 
we also used a diffuser for obtaining the flats. The diffuser plate can be inserted into the 
light path whenever it is necessary. The diffuser was put in front of the objective and the telescope was pointed towards the Sun. Several flat frames were taken to compute the average. 
Typically, the flat fields are taken once a day. If the sky is clear from morning to evening 
then the flats are taken twice a day. The dark exposures are taken as many number of times the flat fields are taken. The average dark exposure was made from the individual dark frames. Later, each individual flat fields are subtracted by the average dark frame. 

Usually, the diffuser produces an intensity pattern which is Gaussian in shape. To remove the gradient in the flat fielded image, a surface fit with a 3rd degree polynomial is used. The resultant fit was subtracted from the flat field. From the residual flat field, the gain table was computed. This is done by dividing the pixel values by the median value of the flat field image. This procedure will rescale the pixel values close to unity. The daily observed solar images are dark subtracted and flat fielded. These flat fielded images are sent for further 
processing before providing it to the scientific community.

\subsection{Image Alignment and Rotation}
Typically, solar images obtained from the telescope will not be in the center of the image window. In addition, for any scientific analysis it is essential to obtain the radius and center of the image. We used the information about the gradient near the solar limb to detect the Sun's limb. Once the solar limb has been detected then we identified the Sun center and radius using the software available in Interactive Data Language (IDL) from Excelis Visual Information
Solutions. This program provides the information about the Sun center and radius with 1-pixel accuracy, which is used to center the Sun in the image window.

Usually, the heliographic north pole will not be along the +y axis of the image in the CCD.
The solar p-angle was calculated for the particular day 
of the month and time by using the standard solar software. The window centered image was 
rotated by the corresponding p-angle in such a way that the solar N-pole along the +y axis 
of the image. All the image alignment programs are written in IDL. The images are stored 
in 32-bit Flexible Image Transport 
(FITS) format with a standard header. We also save the images in JPEG format for quick look purposes.

\section{Representative Results}
tex
\begin{figure}[h]
\begin{center}
   \includegraphics[width=0.5\textwidth]{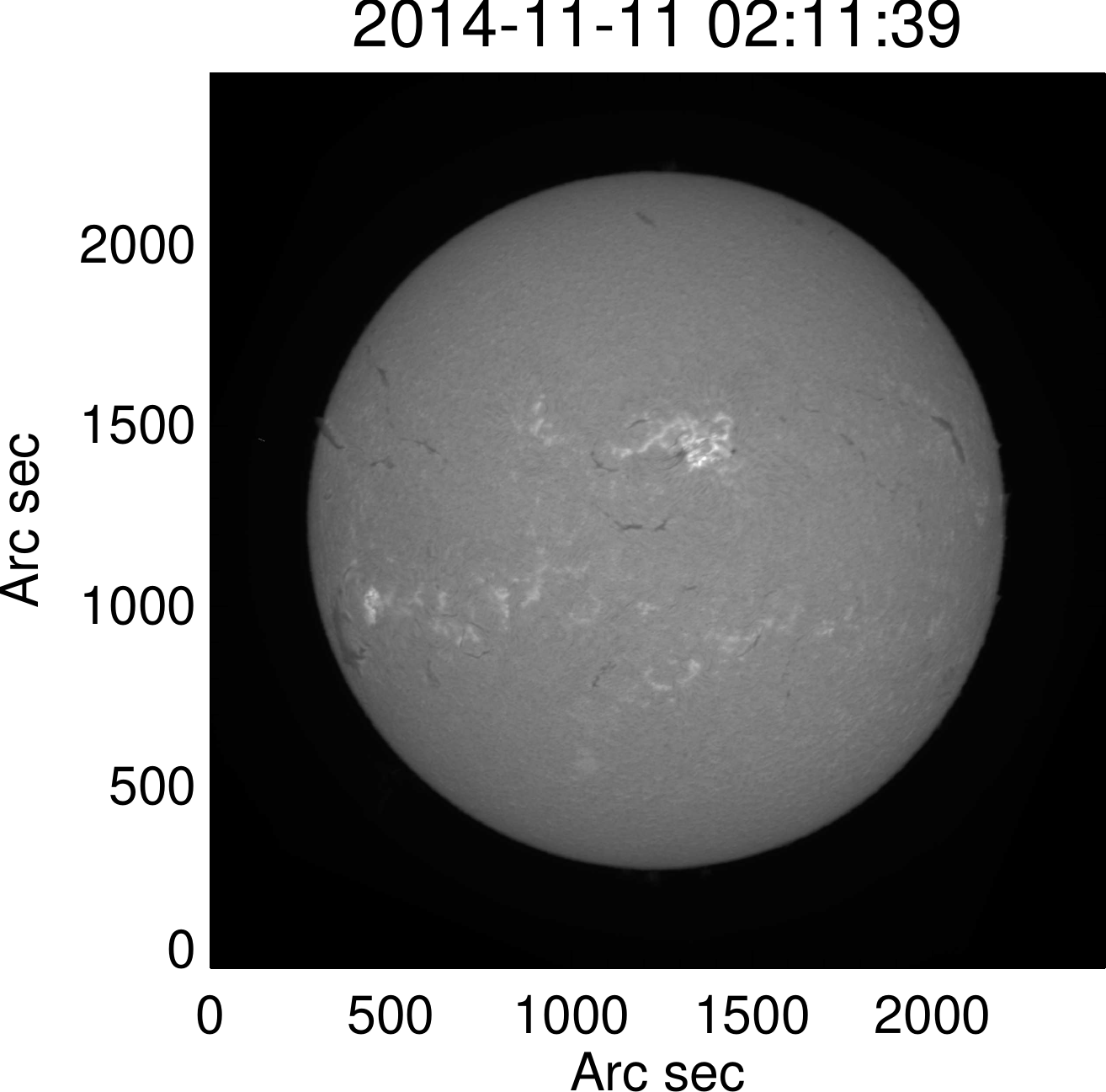}\includegraphics[width=0.5\textwidth]{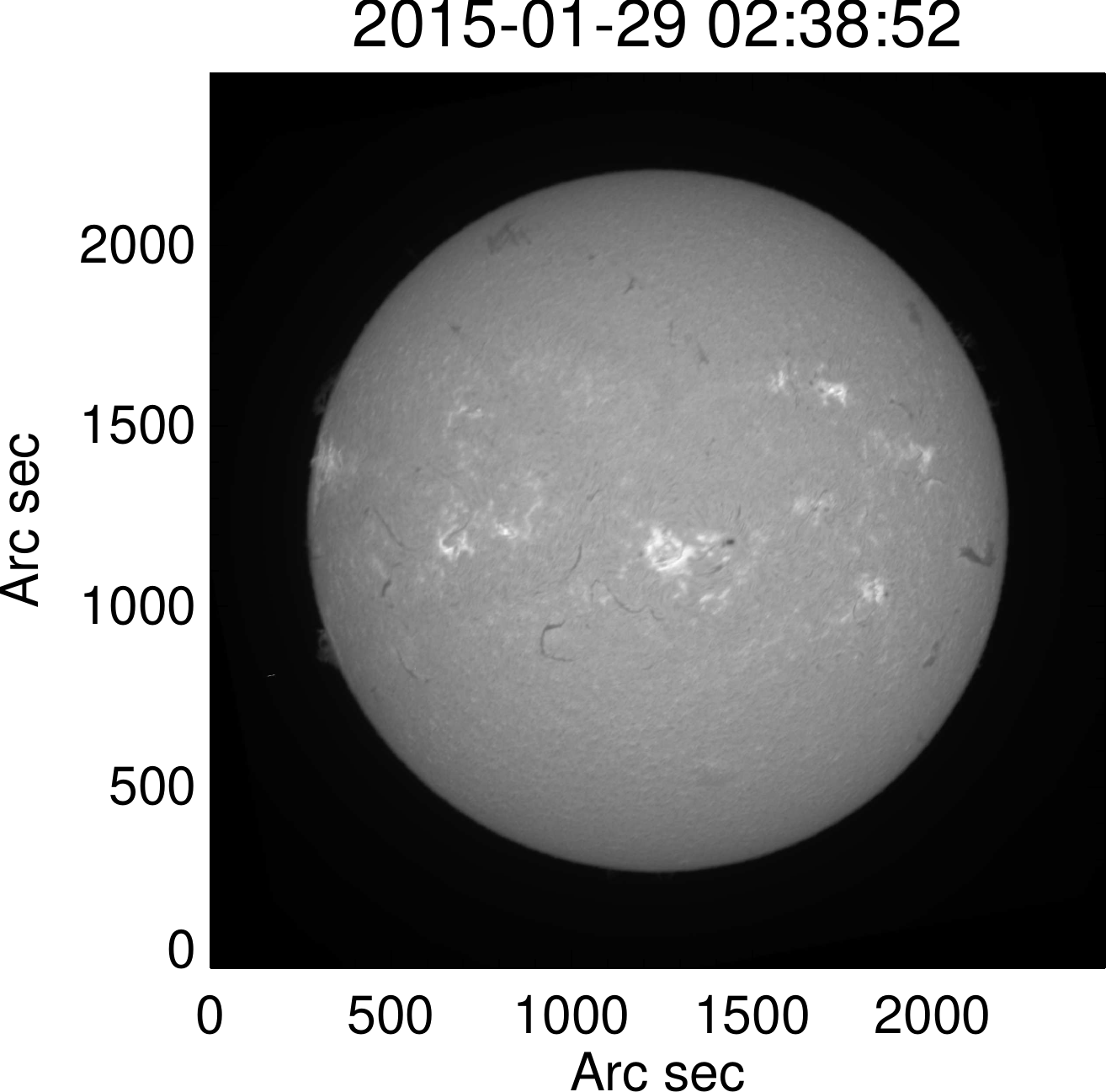}
\end{center}
   \caption{{\small Images taken with the H$_{\alpha}$ telescope installed at the Kodaikanal observatory.}}
\label{fig:3}
\end{figure}

The observations began on Oct. 7, 2014. Since then continuous observations of 
the chromosphere are being carried out with this telescope whenever the sky is clear. 
Figure~\ref{fig:3} shows a couple of images taken in the line center of 
H$_{\alpha}$ on different days. Fibrils, filaments, and plages are clearly seen and the 
image quality is comparable to those from other observatories.

The GONG network telescope started their H$_{\alpha}$ observations in 2010 \citep{Hill2009}. There are 6 stations around the world making  observations of the Sun in this wavelength for a 24~hour coverage. The GONG station at Udaipur Solar Observatory provides the images which are closest in time with the Kodaikanal observatory (KO) images. Hence we 
make the comparison of the Kodaikanal H$_{\alpha}$ images with that of the GONG Udaipur station. The pixel sizes of both the images are different. The GONG pixel resolution is a little higher than the KO images. In order to make a comparison, we rescaled the images to an indentical
pixel size. This was achieved by taking the ratios of the diameters of the full-disk H$_{\alpha}$ images and then decreasing the size of the GONG solar disk in proportion to the 
KO image. 

\begin{figure}[h]
\begin{center}
   \includegraphics[width=0.4\textwidth]{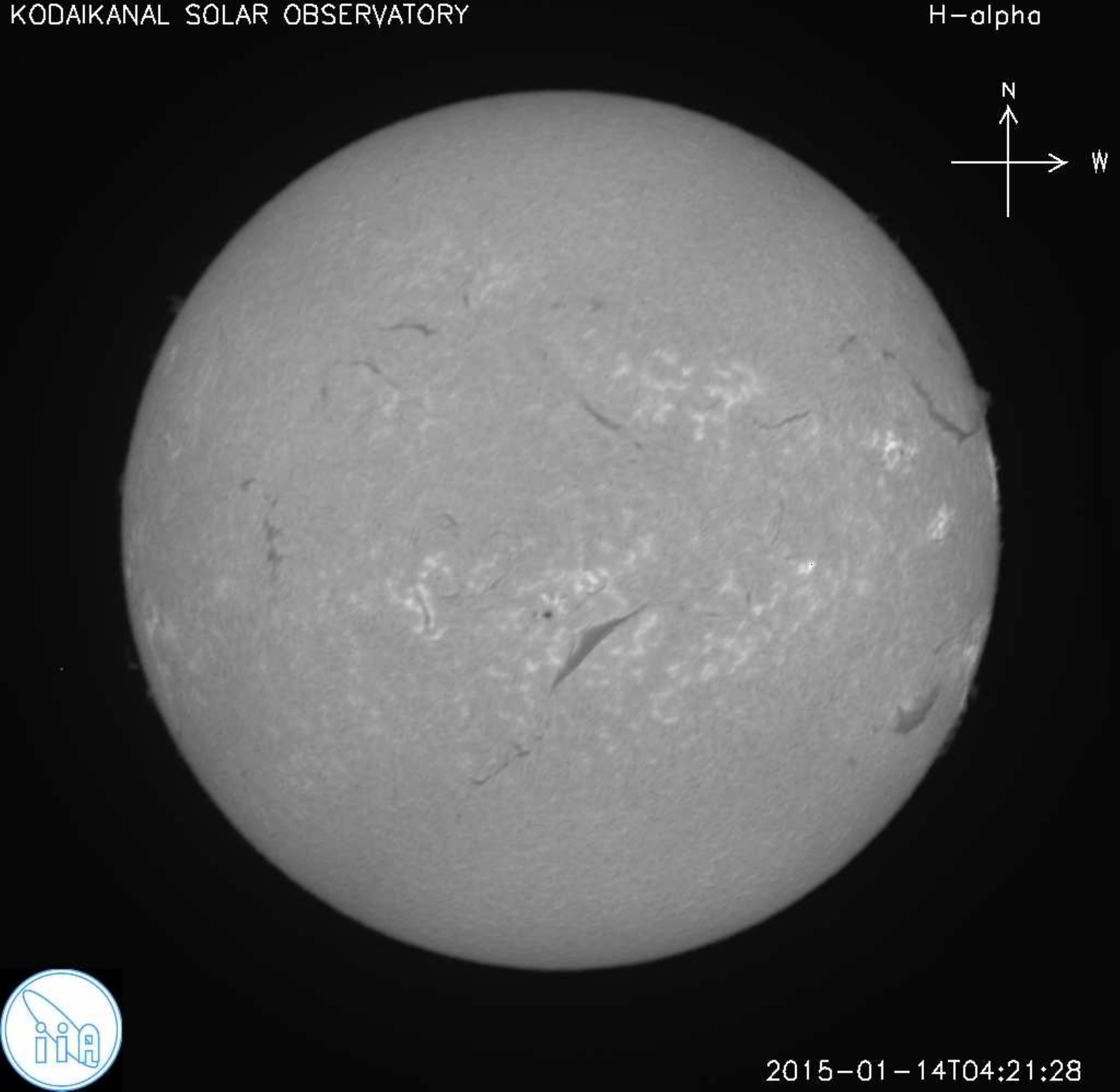}\hspace*{0.05\textwidth}\includegraphics[width=0.4\textwidth]{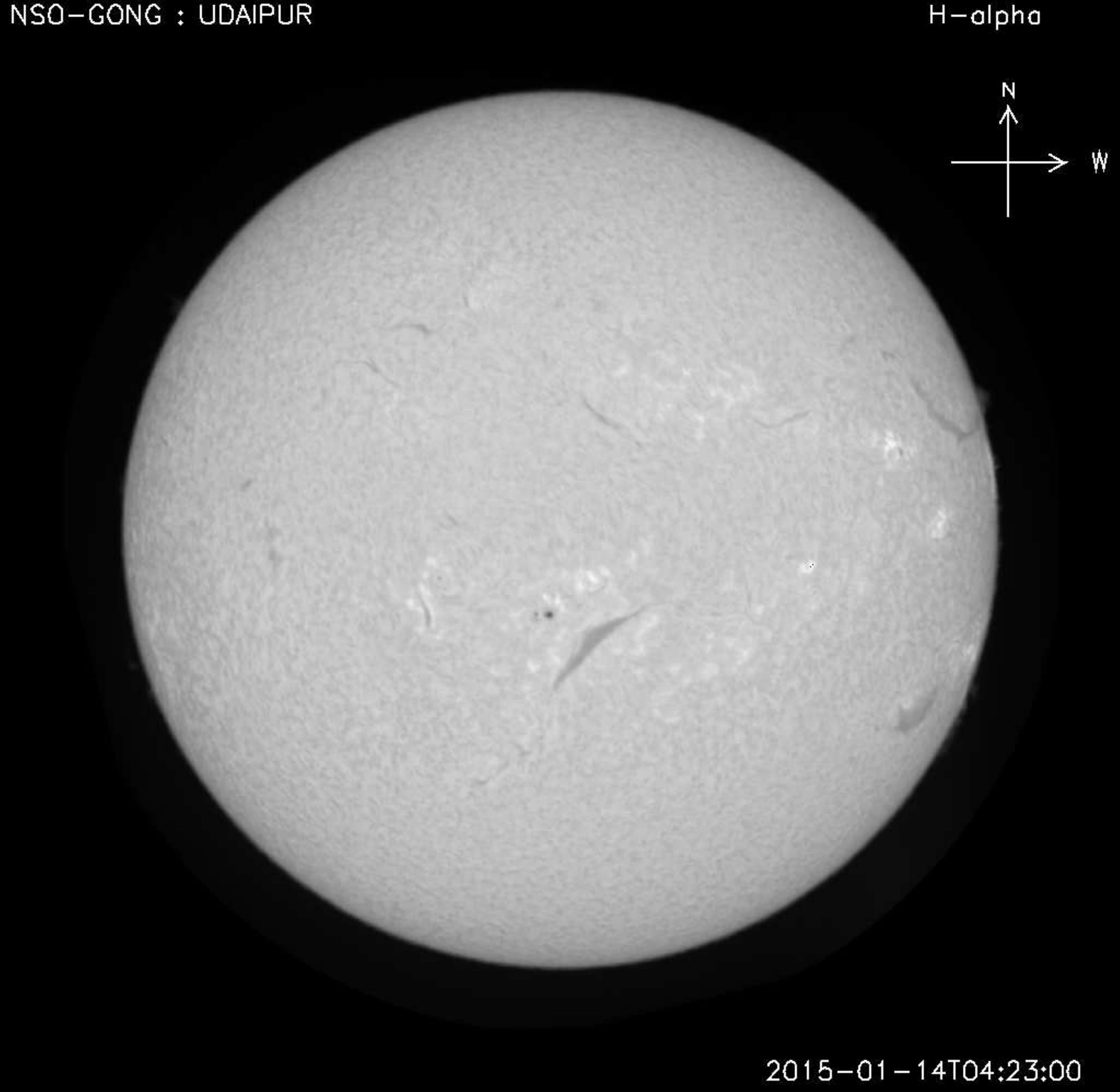}
\end{center}
   \caption{{\small Left: H$_{\alpha}$ image taken with the telescope installed at Kodaikanal observatory. The image was taken on Jan 14, 2015 at 04:21~UT. Right: image taken with 
GONG telescope at the same time.}}
\label{fig:4}
\end{figure}

Figure~\ref{fig:4} shows the KO and GONG H$_{\alpha}$ images side-by-side. The large scale features such as plages and filaments are visible in both the images. The images obtained 
at KO are another useful contribution for H$_{\alpha}$ data coverage.

\begin{figure}[h]
\begin{center}
\includegraphics[width=0.42\textwidth]{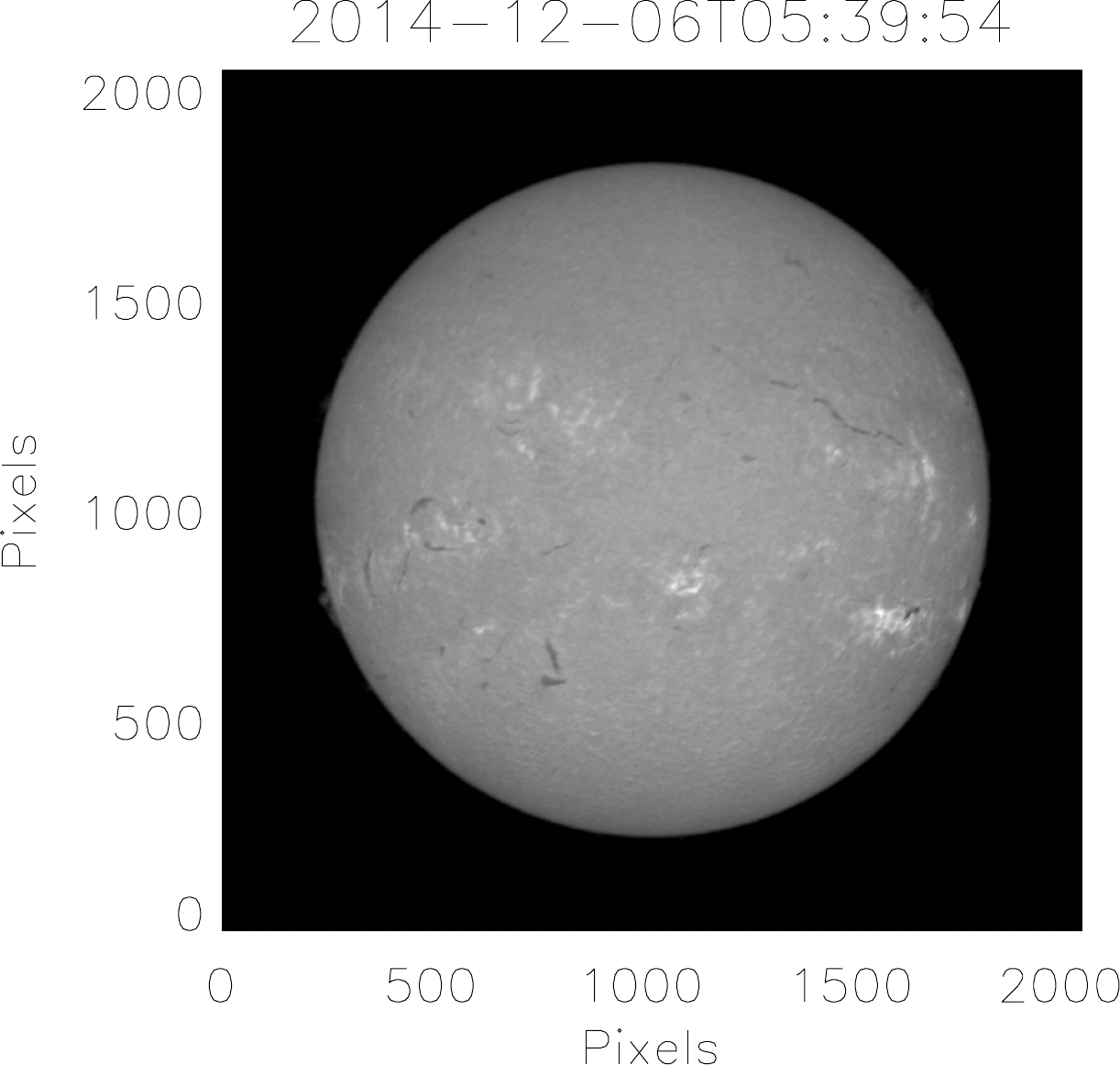}\includegraphics[width=0.58\textwidth]{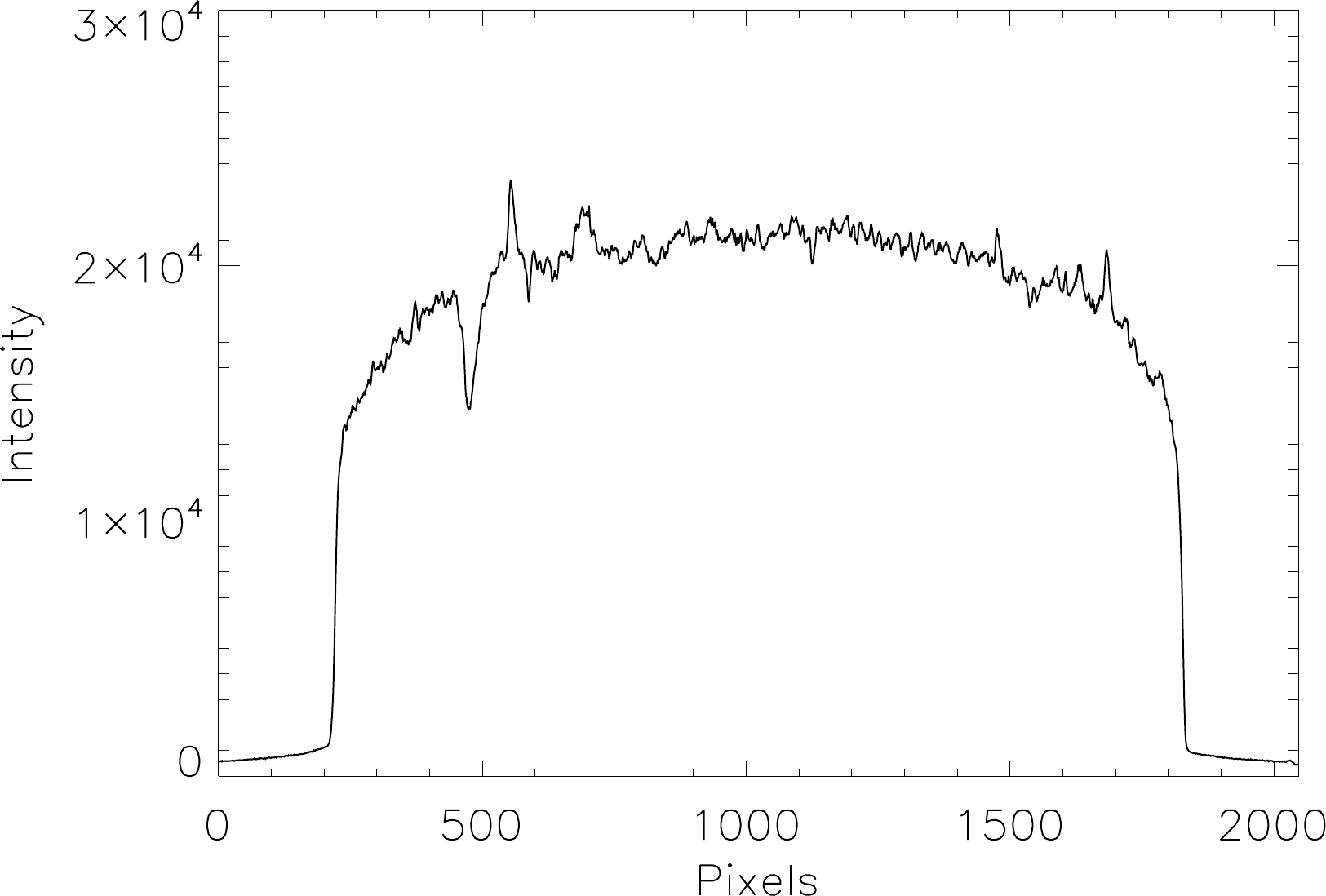} \\
\includegraphics[width=0.42\textwidth]{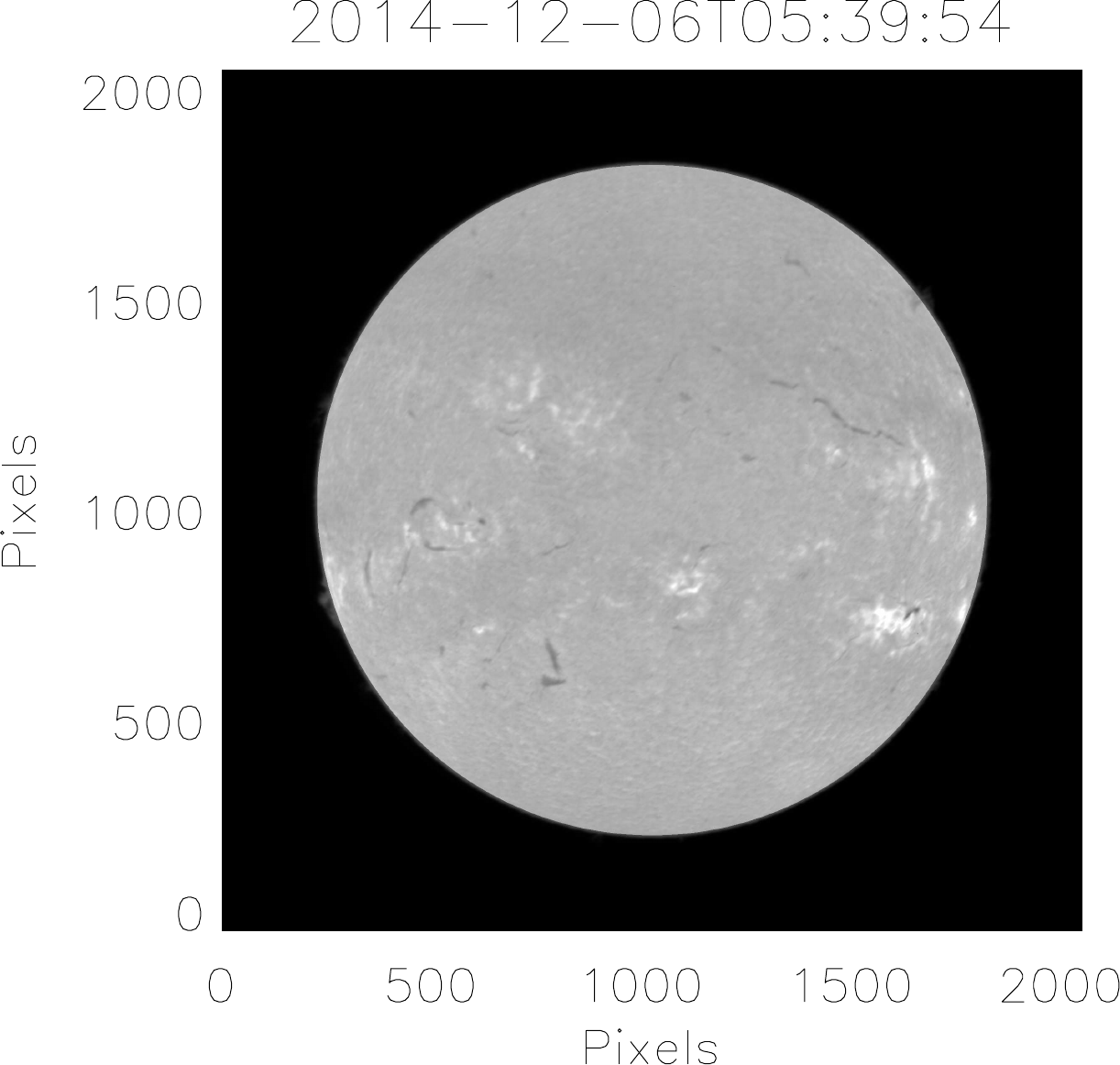}\includegraphics[width=0.58\textwidth]{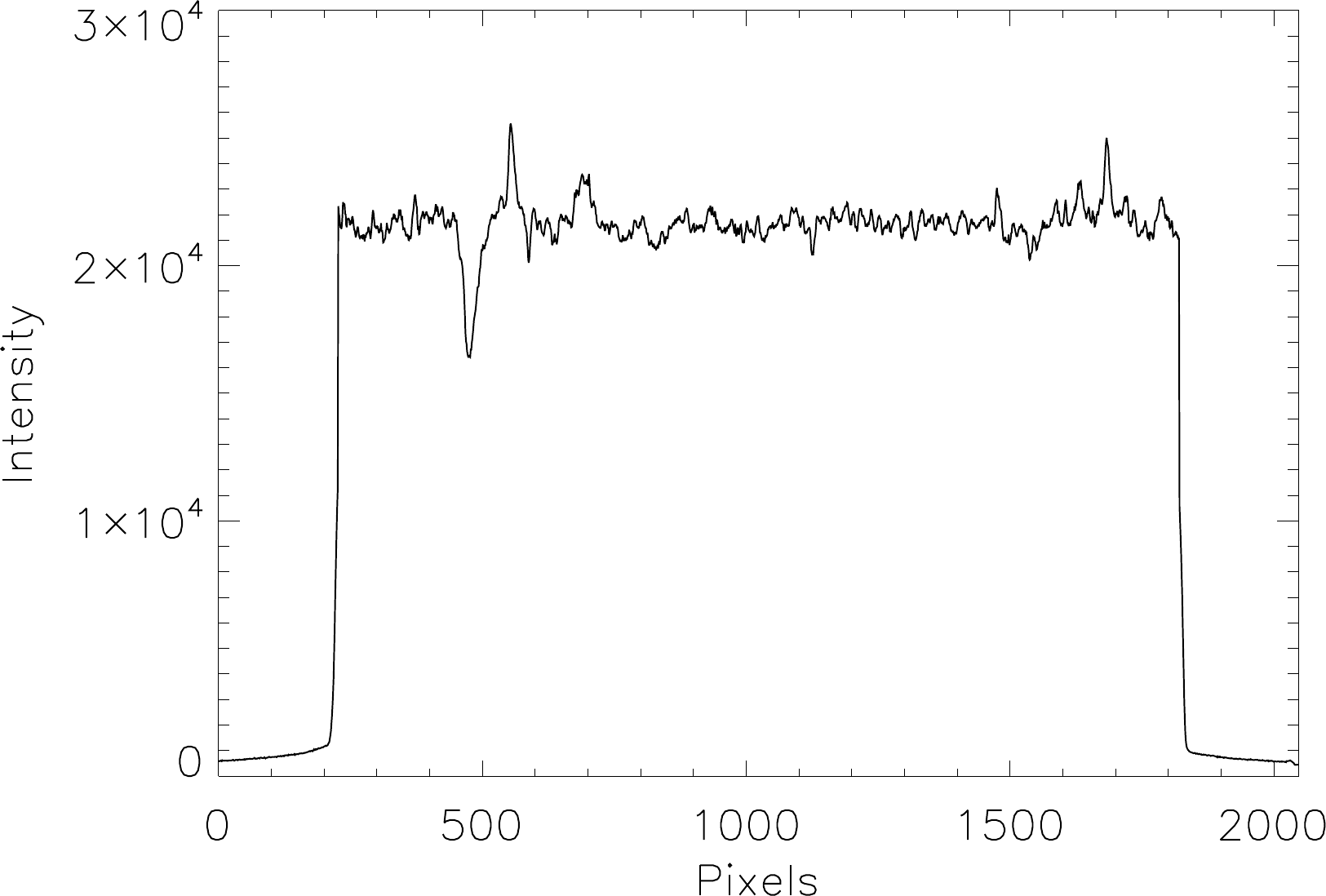} \\
   \end{center}
   \caption{{\small Top-left: H$_{\alpha}$ line center image with center-to-limb variations 
in intensity. Top-right: Limb darkening profile along a row of pixels. Bottom-left:
Same as top-left side image but after removal of center-to-limb variations in intensity.
Bottom-right: Profile after correction.}}
\label{fig:5}
\end{figure}

The line centered H$_{\alpha}$ images show the limb darkening effect (Figure~\ref{fig:5}(top-left)). In order to remove the limb darkening effect we first transformed the image from 
Cartesian to polar coordinates. The average radial profile was constructed by taking the 
median values in the azimuthal direction.
The median value was used instead of mean as plages and filaments are high contrast features 
which could increase or decrease the values.  The resulting profile was transformed to the Cartesian coordinate system and then interpolated to 2-dimensions to fit the solar image. The limb darkening image was normalized to unity and then the image was divided by the limb 
darkening image. This procedure removed the limb darkening effect. In 
Figure~\ref{fig:5} (top-right) we show a plot of the center-to-limb variation of the intensity 
observed in H$_{\alpha}$ image, taken from the central portion of the image (top-left). 
The bottom-right side plot shows the same after the limb darkening correction.
The limb darkening corrected H$_{\alpha}$ image is shown on the bottom-left in 
Figure~\ref{fig:5}. The final image is written in FITS format with the appropriate header information.

 \begin{figure}[h]
 \begin{center}
    \includegraphics[width=0.5\textwidth]{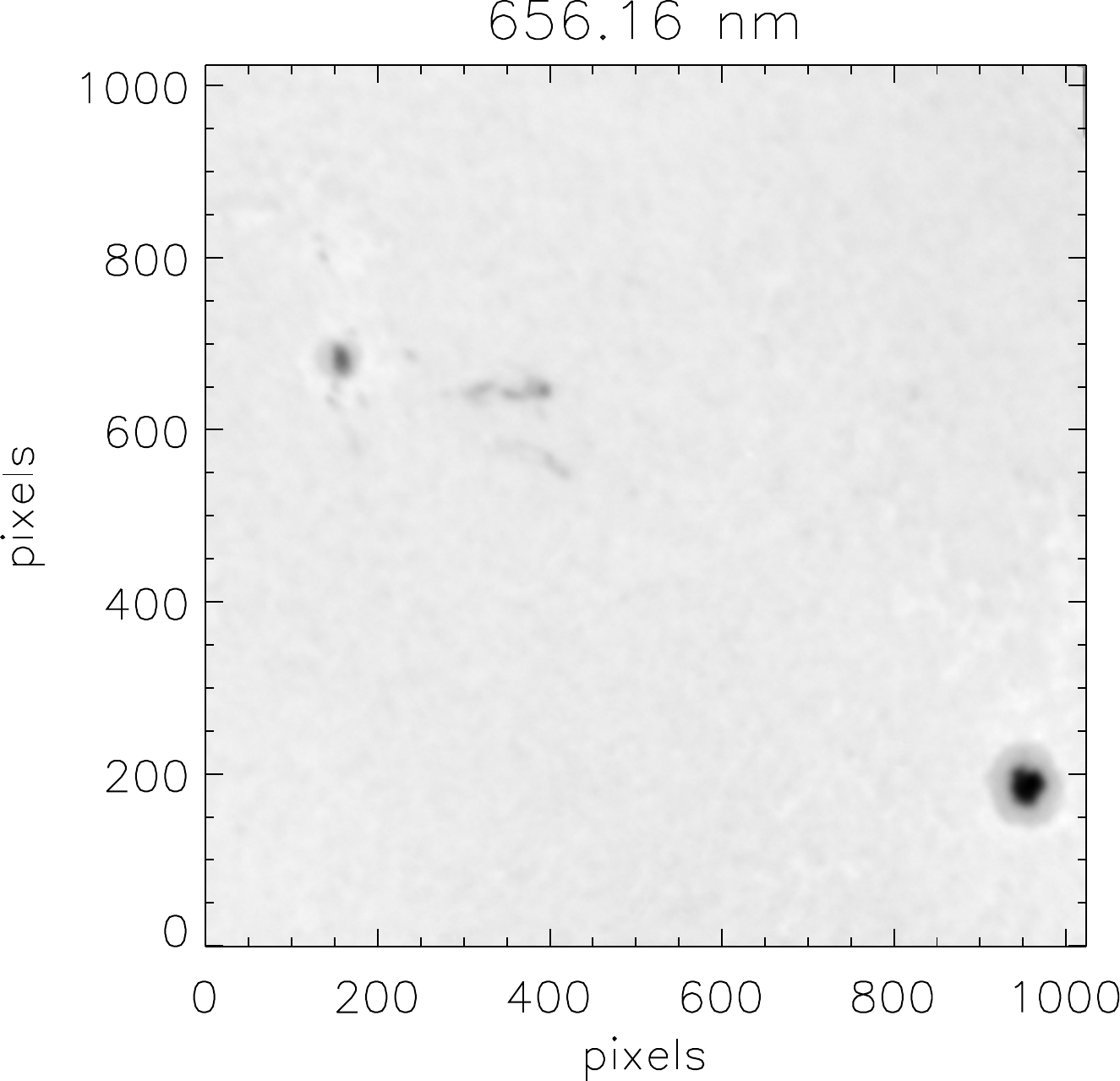}\includegraphics[width=0.5\textwidth]{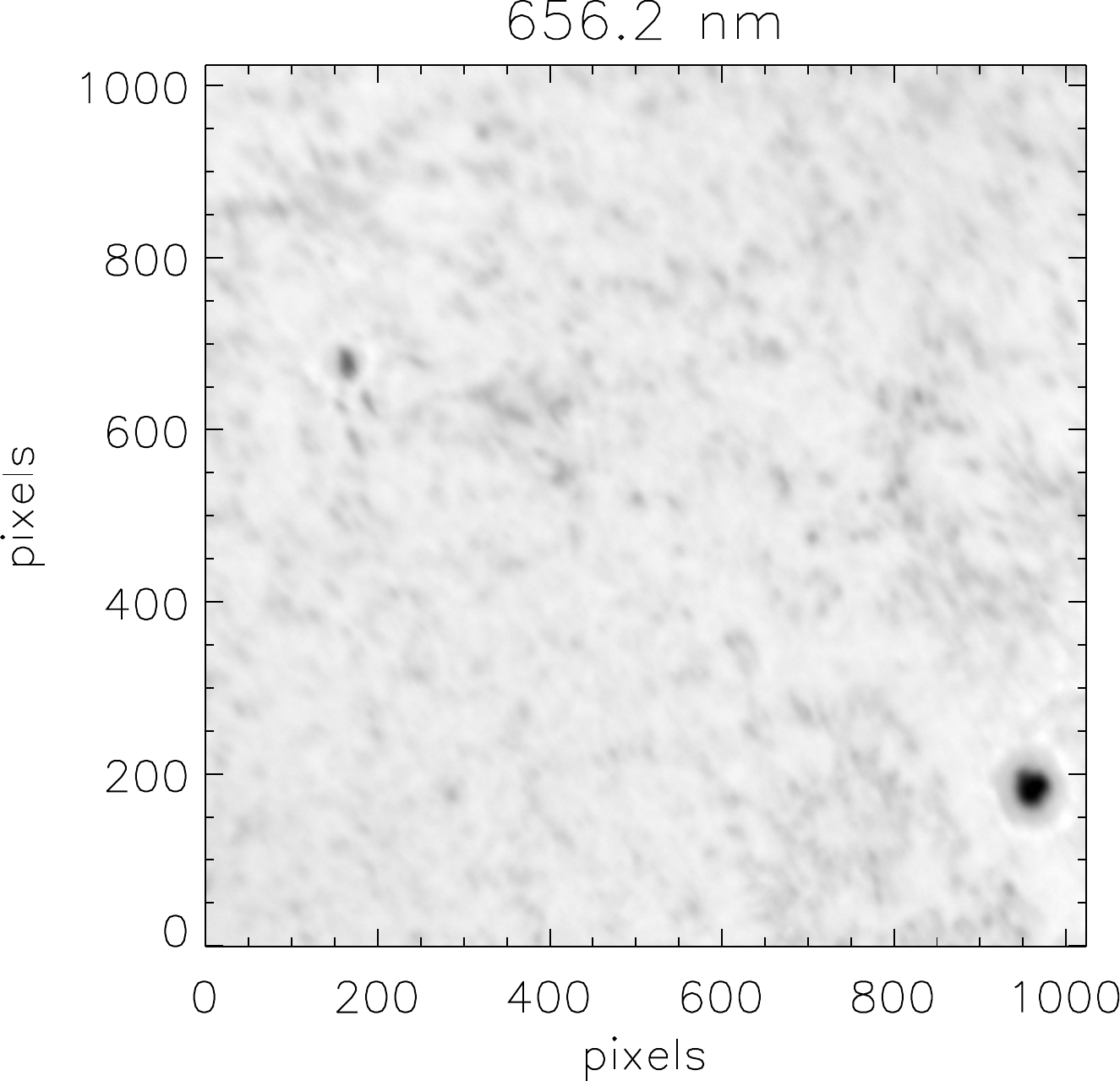} \\
    \includegraphics[width=0.5\textwidth]{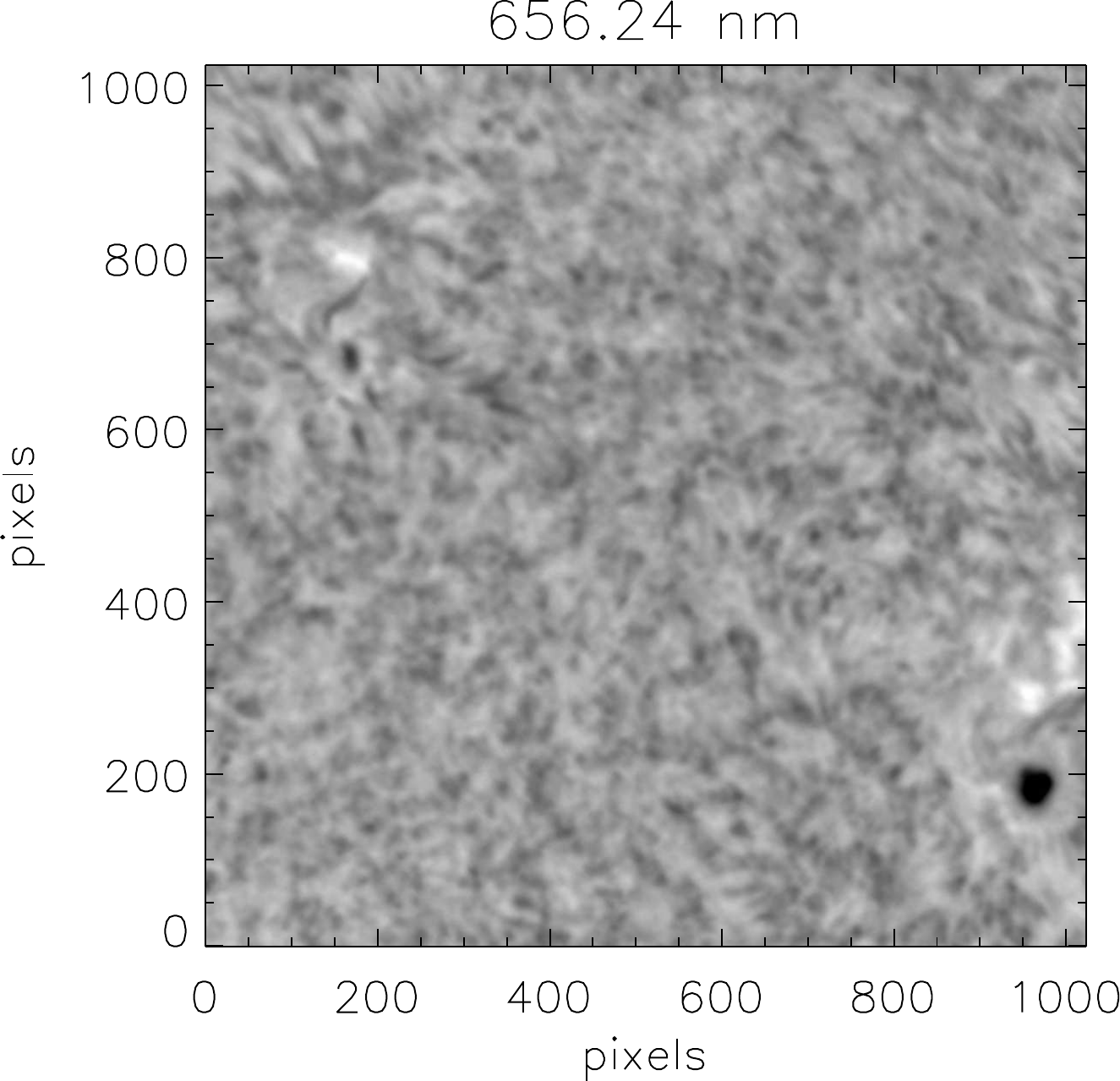}\includegraphics[width=0.5\textwidth]{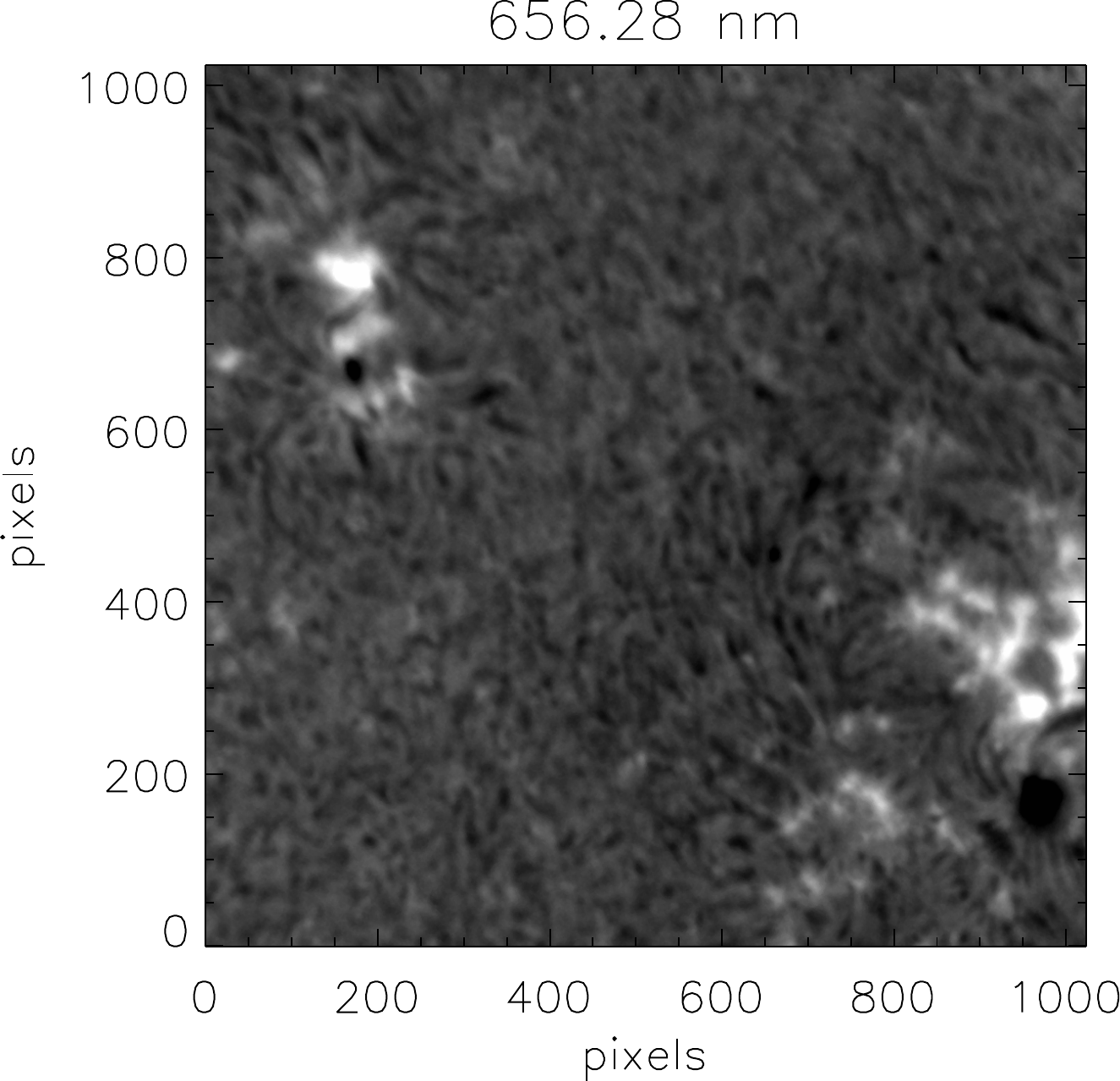} \\
    \end{center}
    \caption{{\small High resolution H$_{\alpha}$ images taken at different positions on the blue wing. Top-left:H$_{\alpha}$ blue wing continuum image (656.28-0.12~nm), top-right:H$_{\alpha}$ blue wing image at 656.20 nm (656.28-0.08~nm), bottom-left:H$_{\alpha}$ blue wing image at 656.24 nm (656.28-0.04~nm) and bottom-right:H$_{\alpha}$ line center image.}} 
 \label{fig:6}
 \end{figure}

\begin{figure}[h]
 \begin{center}
    \includegraphics[width=0.8\textwidth]{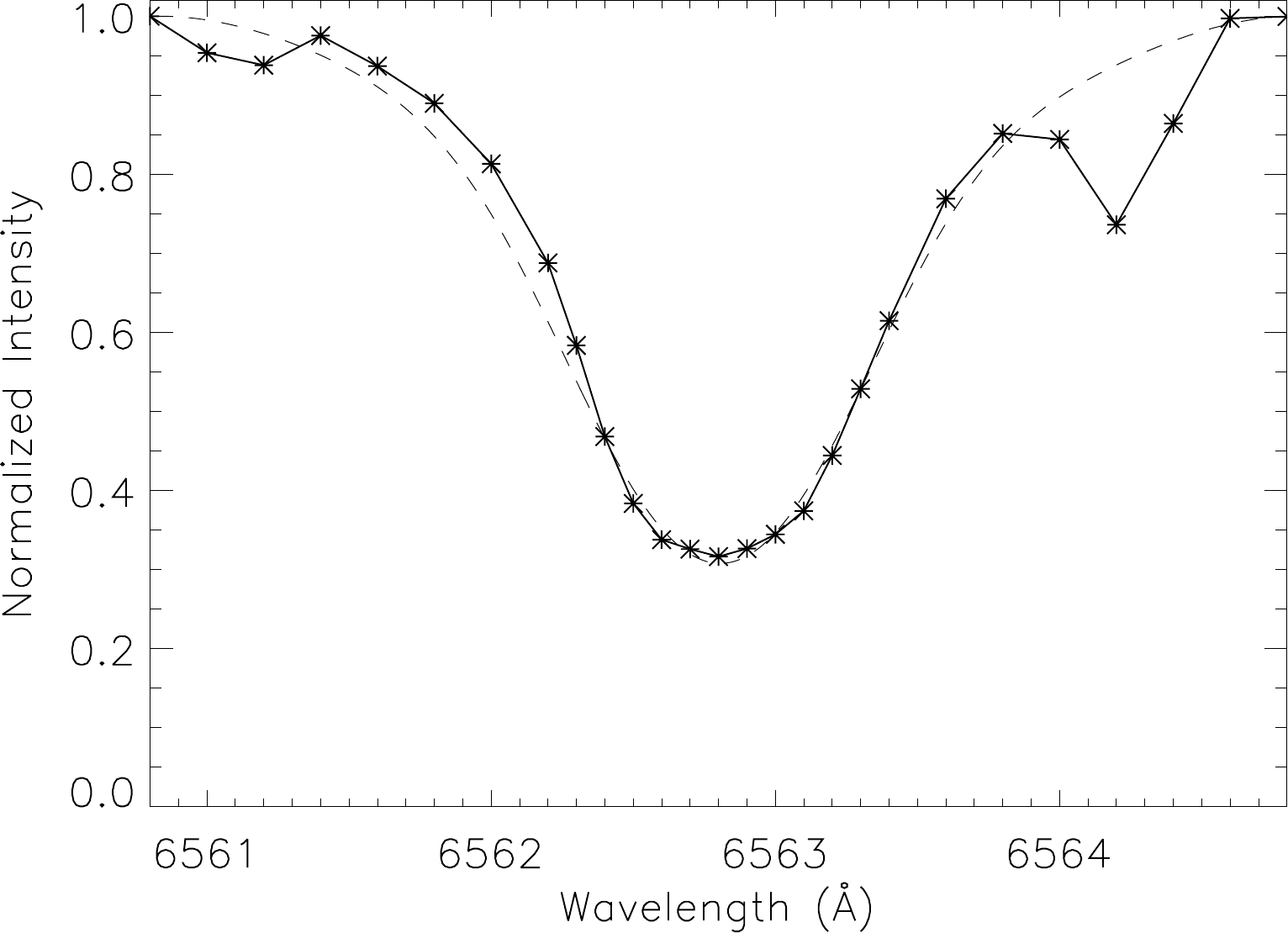}
\end{center}
    \caption{{The reconstructed H$_{\alpha}$ line profile from the set of images taken by 
tuning the H$_{\alpha}$ filter at various positions along the line profile. The asterisks 
symbol denotes the data points obtained from the observations. The dashed line is obtained from 
BASS spectra convolved with the Gaussian kernel having a  width of 0.32\AA.}}
\label{fig:7}
 \end{figure}

We have scanned the H$_{\alpha}$ line profile by taking images at each wavelength position 
at an interval of 0.1~\AA~near the line center and 0.2~\AA~away from the line center. 
These images are obtained by using the relay lens which magnifies the image so that 
1-quarter of the image is covered by the CCD. At each 
wavelength position we have taken the shifted images and using these shifted images we have obtained the flat-field image \citep{chae04}. Figure~\ref{fig:6} shows the images taken at different wavelength positions on the line profile. In each of these images we have taken 
300$\times$300 pixel area in the quiet part of the Sun and then obtained the mean value of 
the intensity. We normalized the intensity to the maximum value in the data points. 
Figure~\ref{fig:7} shows the normalized intensity versus wavelength. A spectra of  
H$_{\alpha}$ line profile obtained from BASS (http://bass2000.obspm.fr/solar$\_$spect.php) 
is convolved with a Gaussian kernel of width ranging from 0.1 to 0.6\AA~and compared 
all the convolved 
spectra with the reconstructed spectra from the observations as has been done in 
\cite{Allende2004}. The comparison shows that 
the BASS H$_{\alpha}$ line convolved with Gaussian of width 0.32\AA~shows better matching 
with the reconstructed profile. This indicates that the passband of the filter is as large 
as 0.32\AA~which is about 0.08~\AA~smaller than measured value in the laboratory. In 
Figure~\ref{fig:7} we show the BASS spectra convolved with a Gaussian kernel of width 
0.32\AA~as a dashed line for comparison.

\begin{figure}[!h]
\begin{center}
\includegraphics[width=0.9\textwidth]{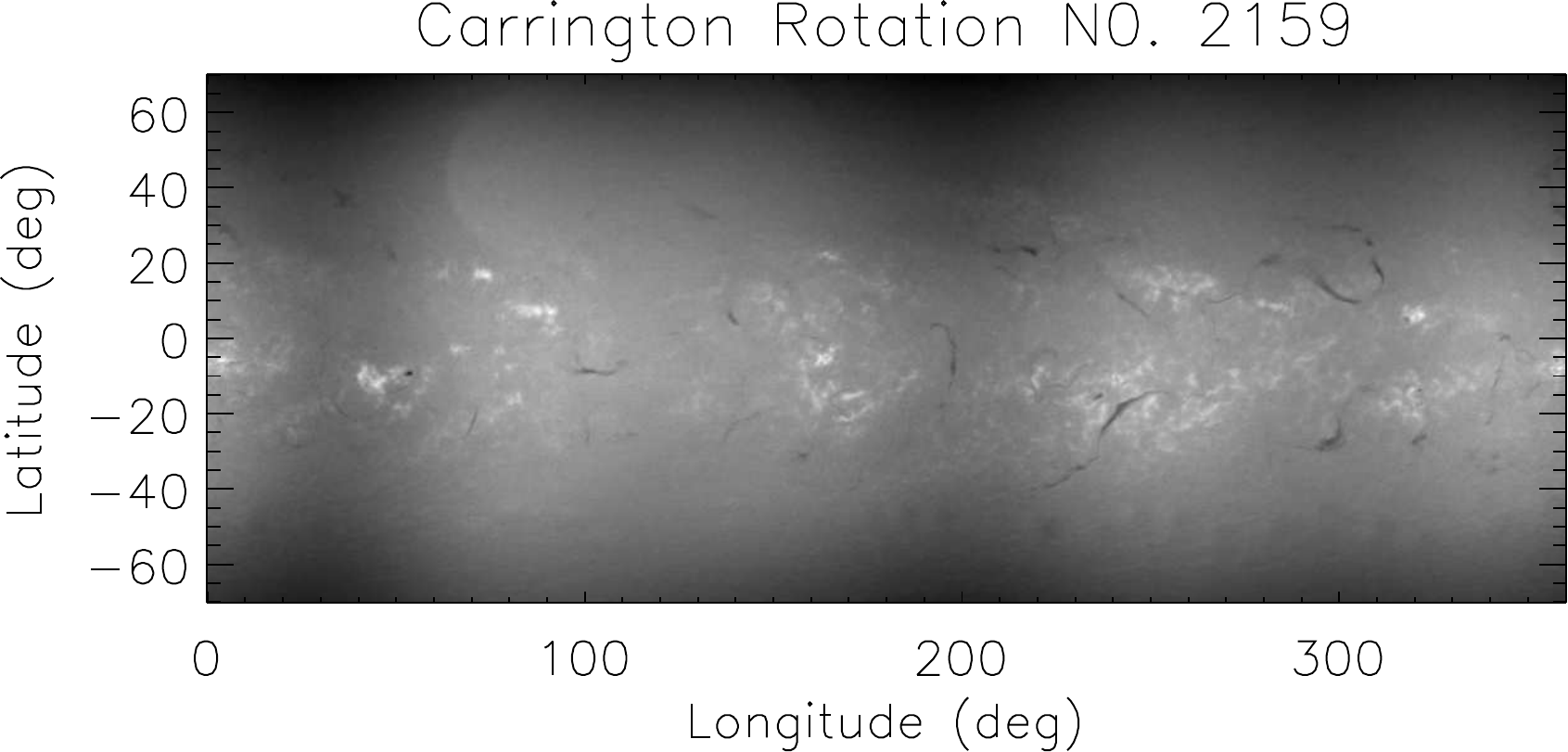}
    \end{center}
   \caption{{\small Chromospheric Carrington map for the Carrington rotation number 2159.}}
\label{fig:8}
\end{figure}

In 1863 Richard Carrington produced synoptic maps for the first time, which provides a 
360$^{\circ}$ view of the Sun in 27 days. The synoptic maps/charts are 27 days of images remapped, put together and plotted in a longitude and latitude co-ordinate system.
Figure~\ref{fig:8} shows the Carrington map for Carrington rotation No. 2159. There were no 
observations on 23rd and 24th January, 2015 but we padded the data gaps with data from 
January 22$^{nd}$. While padding the data we have taken care of the solar rotation.
In the map it is evident that most of the time an intermediate type of filaments exist and   
one can see the active region filaments in only two locations. Plages with and without 
sunspots are also noticeable in the global view. During 2015,  the 
southern hemisphere was predominantly active and plages  appeared between the 
10-15$^{\circ}$ latitude belt.

 \begin{figure}[!h]
\begin{center}
\includegraphics[width=0.5\textwidth]{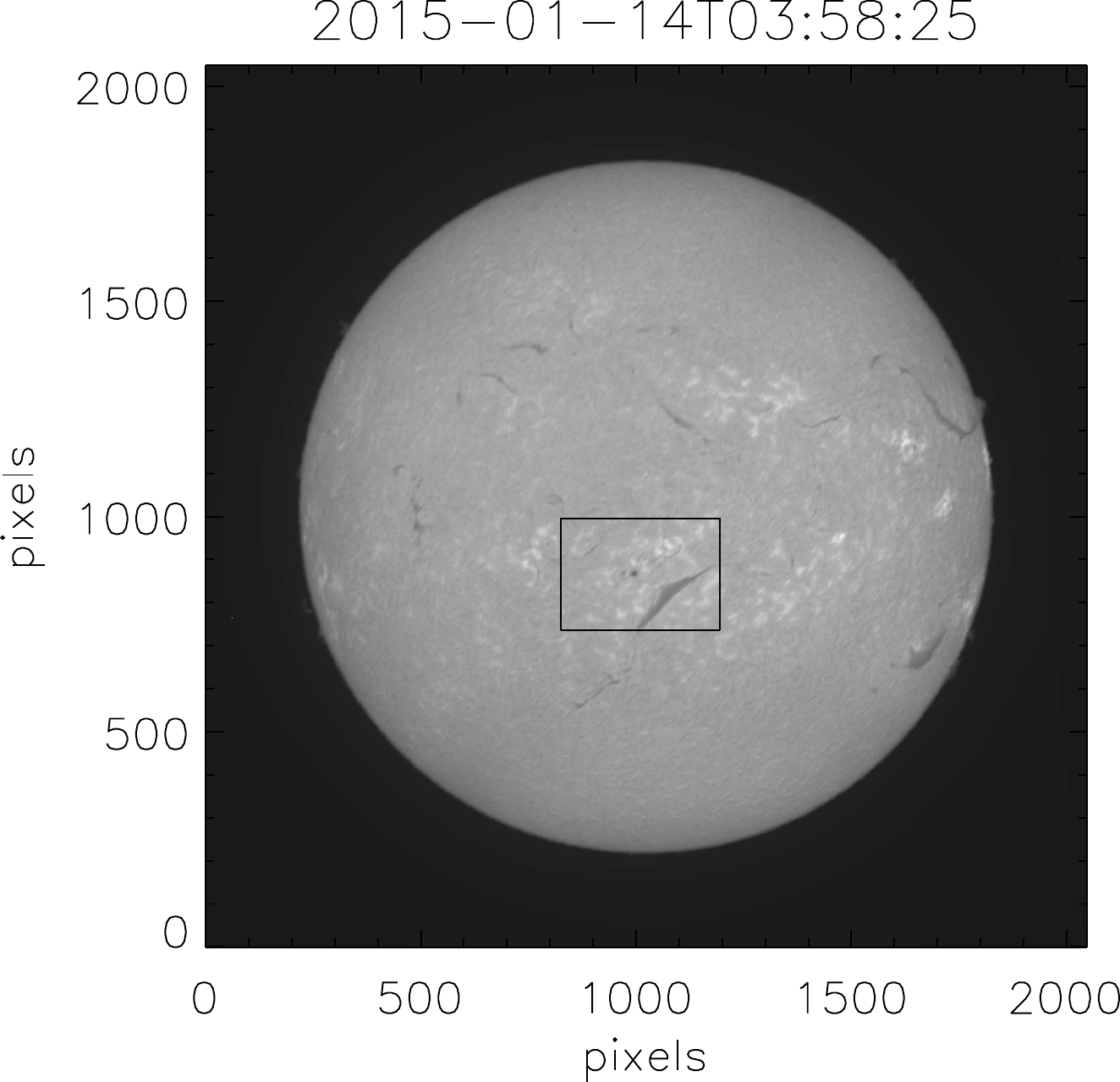} \\
   \includegraphics[width=0.5\textwidth]{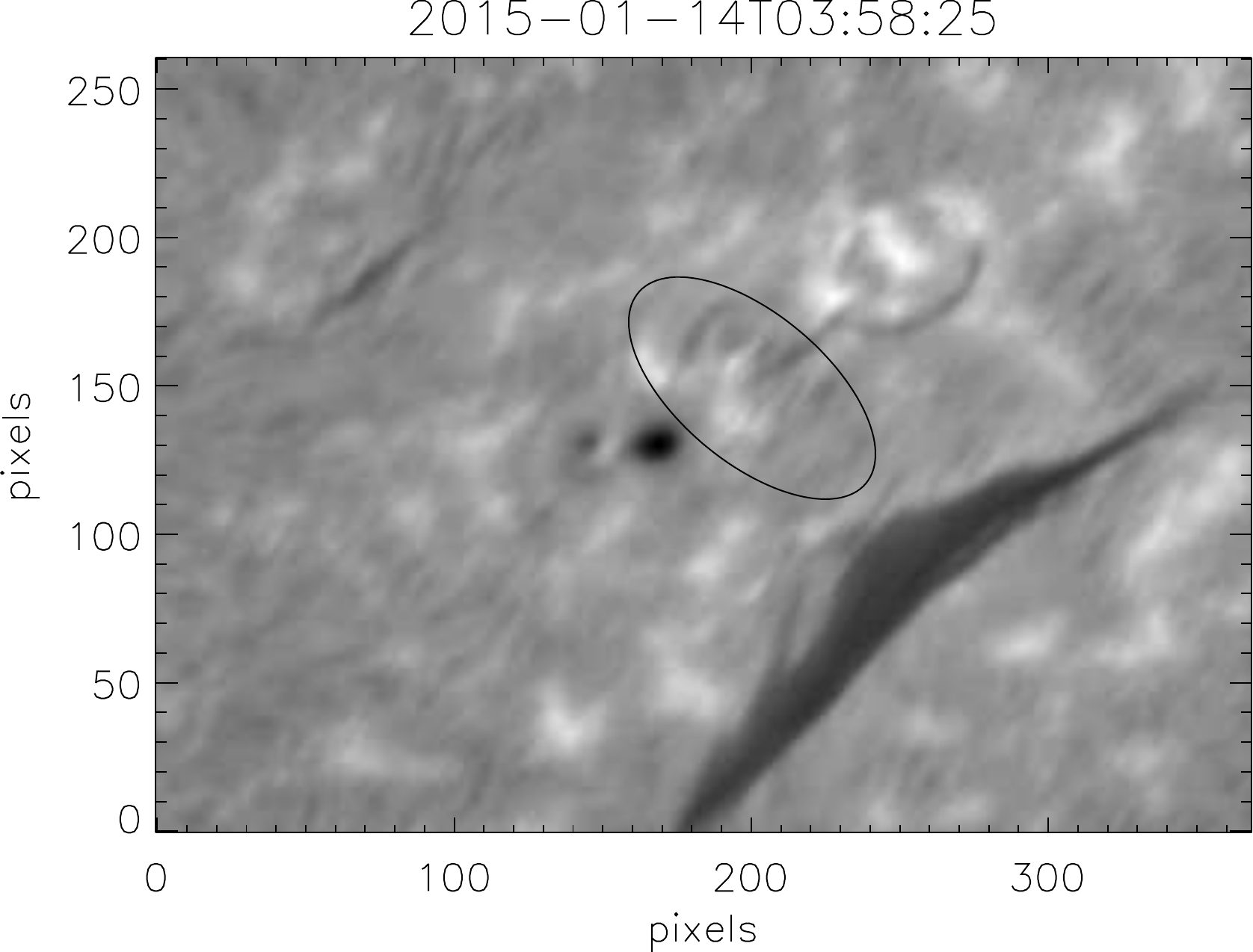}\includegraphics[width=0.5\textwidth]{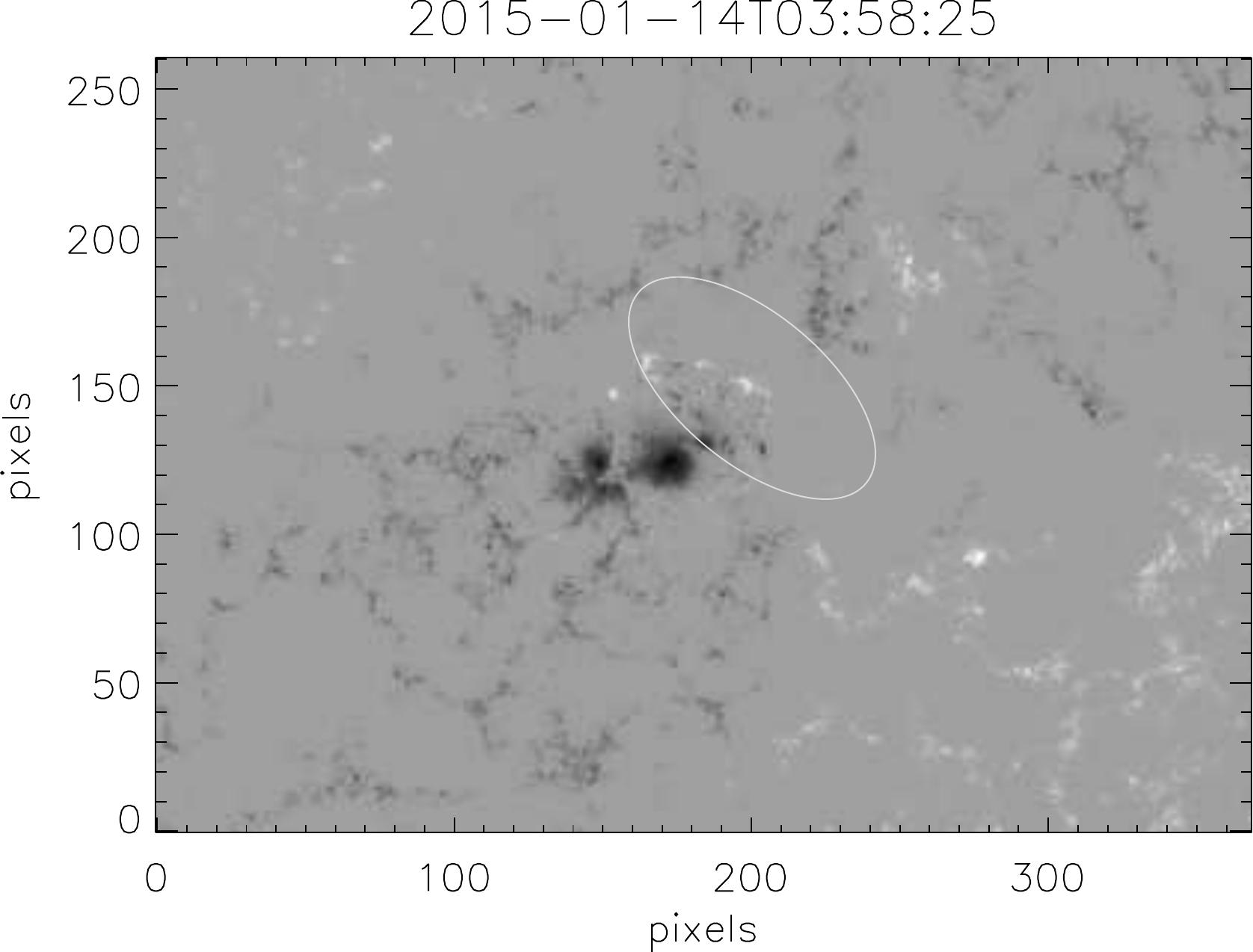} \\
   \end{center}
   \caption{{\small Top: The H$_{\alpha}$ line center image taken on 14 January, 2015. The box represents a field-of-view shown in the bottom-left panel. The corresponding HMI line-of-sight magnetogram is shown in the bottom-right panel. The ellipse in the image shows the region of interest (see text for details).}}
 \label{fig:9}
 \end{figure}

Now, we would like to show one of the preliminary results obtained with this telescope observations in the H$_{\alpha}$ wavelength. Active region NOAA 12259 was a moderate sized sunspot group that appeared on the Eastern limb of the Sun on Jan 08, 2015. On Jan 14, 2015 there were two large and several small sunspots present in this active region. In the southern half of this active region there was a small opposite polarity sunspot. This was surrounded by a large 
area of opposite polarity plage
regions. A filament was formed in this plage region. Figure~\ref{fig:9} (top) shows the 
H$_{\alpha}$ image in full-disk with the sunspot group and filaments. 
A magnified view of the active region in H$_{\alpha}$ wavelength along with a magnetogram 
of the same region taken from the Heliospheric Magnetic Imager \citep[HMI;]{scherrer2012} are also shown in the same Figure (bottom panel). In the H$_{\alpha}$ images, we observed frequent dark and bright ejecta at regular intervals over a period of 
6~hours near the active region (in the location of ellipse). A more detailed study of this phenomenon is required for determining the underlying physical mechanisms.

\section{Summary}
\label{sect:discussion}
Since 1905, Kodaikanal Observatory has a history of observing the Sun in white-light, Ca-K and 
H$_{\alpha}$ wavelengths to monitor solar activity.  We have designed and installed 
an H$_{\alpha}$ telescope for regular observations of the Sun, which has been operational
since October 2014. This telescope is another addition to the existing twin-telescope that
provides full-disk images in Ca-K and white light.

The tunable Lyot filter allows observations at different wavelength positions of the 
H$_{\alpha}$ line. Such a system can be used to construct dopplergrams which we intend
to provide along with full-disk H$_{\alpha}$ images. This telescope is a valuable addition
to the existing facilities designed to continuously monitor the Sun. The data is intended for 
public access and will be made available on a dedicated archive.

\normalem
\begin{acknowledgements}
The telescope installed Kodaikanal observatory is designed and tested for its performance 
at Nanjing by Jagdev Singh (IIA) along with Nanjing Institute for Astronomical Optics \& Technology (NIAOT), CAS, China. We thank the telescope team at NIAOT for their help and support. 
We thank all the members of the telescope installation team: Weijun Mao, Junping Zhang, Haitain Lu, Qiqian Hu, Houkun Ni, Xiaojun Zhou, Qingsheng Zhu, Guilin Wang, Chuanmin Li, \& Xianhua Han for putting the special effort to come to India and install the telescope at the Kodaikanal observatory. We also would like to thank the team members at IIA for their help during the installation of the telescope:  Ananth, A. V., Tsewang, Dorjai, Sagayanathan, K, Nagaraj, B. and 
Mallappa, D., Ismail, J. R., Robert, V., Ravi, K., Rajalingam, Manoharan, J, Fayaz \& Ashok. We also would like to thank Mr. Murali Das for photgraphing the installation event. We also would like to thank the local administrative and technical staff at the Kodaikanal observatory:   Kumaravel, P., George, F., Basha, M. I., Hariharan, G., Devendran, P., Michael, P., Loganathan, D. With out their help it was not possible to install the telescope during those difficult conditions. Thanks also to the director IIA, Sreekumar for his 
constant support and encouragement during the installation of the telescope. We  
thank the referee for fruitful comments which helped us to improve the manuscript.

\end{acknowledgements}

%\bibliographystyle{raa}
%\bibliography{bibtex}

\end{document}